\documentclass[12pt, a4paper, twocolumn]{article}

\usepackage{graphicx}
\usepackage[dvipsnames]{xcolor}
\usepackage{amsmath,amsfonts,bm,amssymb}
\usepackage{comment}

\usepackage[a4paper, total={7in, 9in}]{geometry}
\usepackage[caption=false,font=normalsize,labelfont=sf,textfont=sf]{subfig}
\usepackage{textcomp}
\usepackage[style=ieee, citestyle=numeric-comp, backend=biber]{biblatex}

\usepackage{abstract}

\usepackage{titlesec}
\titleformat*{\section}{\large\bfseries}

\begin{document}

\title{Numerical simulation of the electromagnetic wave reflection from 2D random semi-infinite strongly scattering media}
\author{Sofia Ponomareva, and Alexey A. Shcherbakov \\ {\it School of Physics and Engineering, ITMO University, St-Petersburg, Russia
}}
\date{}
\maketitle

\begin{abstract}
Light scattering in disordered media plays an important role in various areas of applied science from biophysics to astronomy. In this paper we study two approaches to calculate scattering properties of semi-infinite densely packed media with high contrast and wavelength scale inhomogeneities by combining the Fourier Modal Method and the super-cell approach. Our work reveals capabilities to attain ensemble averaged solutions for the Maxwell's equations in complex media, and demonstrated numerical convergence supports the consistency of the considered approaches.
\end{abstract}


\section{Introduction}
\label{sec:intro}

Simulation of the electromagnetic scattering from random media is of considerable interest in many areas of physics. For example, models of light transport in tissues are required for non-invasive diagnostics in the near-infrared and visible spectral bands \cite{Gibson2005Feb}, yielding information about optical and physiological parameters of biological materials. Optical response measurement is one of the main tools of remote sensing and geoscience \cite{Tsang2017Jul} where researchers often deal with rough surfaces of densely packed substances. Study of light propagation in artificially disordered structures is also of great interest, since it allows the engineering of composite materials with specific properties \cite{Yu2021Mar} or exploit specific techniques for imaging through random media \cite{Yoon2020Mar}. However, examination of heterogeneous macroscopic objects remains a challenging computational problem. Especially, when it comes to a strong-scattering regime, in which multi-scattering effects can not be disregarded, like in certain natural materials \cite{Jacucci2020Jul}.

Various approaches can be used to address the problem of light propagation in random media. Firstly, there are methods based on the radiative transfer theory, which is aimed at solving an equation for intensity of a light beam propagating through the media \cite{Mishchenko2003}. It is best suited for a description of sparse media, with relatively low density of scatterers. Some modifications were proposed to expand a range of its applicability to denser particle arrangements \cite{Tishkovets2006Sep}. For densely packed strongly scattering systems one needs accurate solutions of the macroscopic Maxwell equations. Widely used numerical simulation techniques include the finite-difference time-domain method, the finite element method, the discrete dipole approximation, and the superposition T matrix approach \cite{Kahnert2016Jul}. They provide numerically rigorous description of wave propagation in media and are applicable for scatterers of various shapes. One of the major problems of these approaches consists in the dramatic increase of required computational resources with an increase of the computational domain, particle size, and number of scatterers. Although the finite-element, discrete dipole and finite-difference methods can in principle handle general electromagnetic problems with arbitrary material boundaries, such methods do not scale well to the dimensions relevant to microscopic samples. Therefore, there is a general demand to decrease computational costs. Promising candidates are variations of the fast superposition T-matrix methods capable to address ensembles comprising up to $10^5$ particles \cite{Markkanen2017Mar}, \cite{Mackowski2011Sep} and hybrid radiation transfer methods \cite{Roger2014}. With such an approach the authors of \cite{Penttila2021Mar} demonstrated that starting from $10^7$ particles backscattering efficiency of media consisting of randomly arranged spheres converges. An important alternative, which in principle should allow one to perform benchmarking with these powerful methods when tracing a passage to infinite media, is a super-cell approach for methods adapted to simulation of periodic systems. An example is a recent study \cite{Theobald2021Sep}, which demonstrated a combination of the superposition T-matrix method with lattice summation to describe light scattering from porous polymer layers by periodic particle arrangements. To mention here, apart from rigorously solving electromagnetic equations, there was proposed another approach consisting in generation of random matrices describing scattering from disordered media \cite{Byrnes2022}. 

In this work we further explore the supercell approach, and in contrast to the mentioned study we adapt the Fourier modal method (FMM) instead of the periodic superposition T-matrix method. The latter one is a well established and versatile tool for computing periodic diffractive structures \cite{Gushchin2010}. An additional advantage in application of the FMM is that the plane wave decomposition being intrinsic to the method naturally allows to consider both semi-infinite and bounded scattering layers inside arbitrary planar mulatilayer structures, as shown below.


\section{Fourier Modal Method for super-cells}
\label{sec:fmm}
The super-cell approach is one of the methods to simulate disorder or randomness in large scale or complex physical systems under limited computing resources. Its consistency was already demonstrated in quantum mechanics, quantum chemisty, solid state physics, as well as in electromagetics (e.g., \cite{Theobald2021Sep,Jarvis1997,Dovesi1994,Elson2003,Bulgakov2000}). Here we are aimed to apply the super-cell approach to 2D random strongly scattering structures by introducing an artificial periodicity with period $\Lambda$, and study the numerical performance of the overall method.

\begin{figure}[ht]
    \centering
\includegraphics[width=0.5\textwidth]{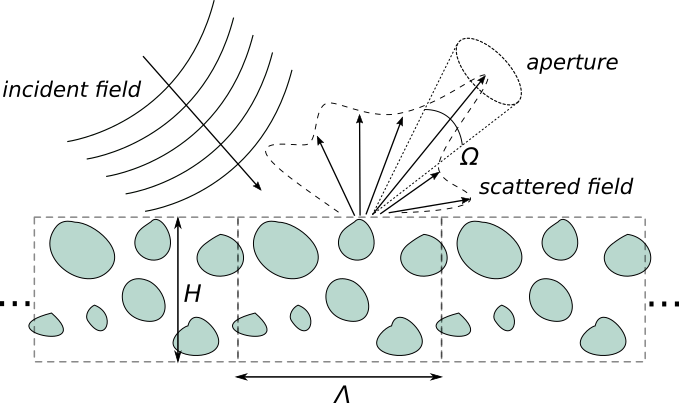}
\caption{Illustration of the scattering problem and the super-cell approach.}
\label{fig:ill_system}
\end{figure}

Taking into account the introduced periodicity we use the Fourier Modal Method (FMM) as a cornerstone of our approaches due to the widespread and popularity of this method for analysis of periodic electromagnetic structures. The FMM is designed to solve the time-harmonic Maxwell's equations for planar slabs of 1D and 2D photonic crystals in order to calculate their scattering properties. It consists of the following three major steps. Firstly, one solves an eigenvalue problem for a given infinite photonic crystal to calculate mode propagation constants and Fourier amplitude vectors of mode fields. At the second step the calculated quantities are used to find T-matrices of interfaces separating the slab and a surrounding media. On the third step one combines T-matrix blocks to yield slab scattering matrix (S-matrix).

Given a periodic slab parallel to $XY$ plane of Cartesian coordinates $XYZ$ with periodicity along $X$ axis, let us denote vectors of Fourier amplitudes of incoming and outgoing waves, \emph{a}bove and \emph{b}elow the slab, as $\bm{a}^{\pm}$, and $\bm{b}^{\pm}$. The sign "$\pm$" indicates propagation direction relative to axis $Z$. Each component of the vectors is amplitude of a corresponding plane wave, e.g. $a^{\pm}_m\exp(ik_{mx}x\pm ik_{mz}z-i\omega t)$. The in-plane wave vector projections are governed by the grating equation 
\begin{equation}
    k_{mx}=k_b+2\pi m/\Lambda,\;m\in\mathbb{Z},
    \label{eq:grating}
\end{equation}
the vertical projections -- by the dispersion equation
\begin{equation}
    k_{mz}=\sqrt{\omega^2\varepsilon\mu_0-k_{mx}^2}
    \label{eq:dispersion}
\end{equation}
with $\varepsilon$ being permittivity of a medium either below or above the slab. With these notations the FMM allows to compute an S-matrix, which can be represented though its reflection and transmission blocks:
\begin{equation}
	\left(\!\!\begin{array}{c}\bm{b}^- \\ \bm{a}^+\end{array}\!\!\right) = S \left(\!\!\begin{array}{c}\bm{b}^+ \\ \bm{a}^-\end{array}\!\!\right) = \left(\!\!\begin{array}{cc} R_{bb} & T_{ba} \\ T_{ab} & R_{aa} \end{array}\!\!\right) \left(\!\!\begin{array}{c}\bm{b}^+ \\ \bm{a}^-\end{array}\!\!\right).
	\label{eq:smat}
\end{equation}
Given the maximum number of Fourier harmonics $N_h$ taken into account in calculations, the size of S-matrix is $2N_h\times2N_h$.

\begin{figure}[h!]
    \centering
\includegraphics[width=0.35\textwidth]{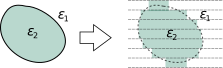} \caption{Illustration of the slicing approach.}
\label{fig:ill_slicing}
\end{figure}

A 2D scattering medium is described by a spatially varying permittivity $\varepsilon(\bm{r})$, which is a $\Lambda$-periodic function of the coordinate $X$ within the super-cell approach, as illustrated in Fig.~\ref{fig:ill_system}. The scattering medium is described by a characteristic geometry of inclusions, their packing density and average size. Using these parameters we generate particular samples of scattering structures adding the artificial periodicity. The medium is supposed to be infinite in the half-space $|z|\leq0$, so that a material "above" it ($z>0$) is homogeneous and isotropic. To apply the FMM for such structures one has to consider, first, finite height of samples (denoted as $H$ in Fig.~\ref{fig:ill_system}), and, second, slicing into plane parallel layers along vertical axis $Z$. This partitioning is illustrated in Fig.~\ref{fig:ill_slicing}. The slice thickness $\Delta h_s$ is small enough to approximate its permittivity distribution by step function and treat each slice as a photonic crystal slab with $z$-independent permittivity. S-matrices of different slices can be then combined into S-matrices of thick (relatively to the wavelength) scattering layers according to specific multiplication rules \cite{Ko1988} (see Appendix).

For analysing reflection properties of semi-infinite structures only $R_{aa}$ S-matrix block is of interest. When multiplied by a Fourier vector of an incident field it yields an amplitude vector of the scattered field (reflected waves). The introduced artificial periodicity imposes equidistant discretization in the $k$-space in accordance with the grating equation, Eq.~(\ref{eq:grating}), which in turn implies a discrete set of accessible propagation angles $\theta_m  \leftrightarrow k_{mx}$  for scattered plane waves. A scattering problem commonly requires evaluation of the power $P_{sca}$ reflected to a given aperture $|\theta-\theta_0|\leq\Delta\theta/2$ as demonstrated in Fig.~\ref{fig:ill_system}. Therefore, to estimate $P_{sca}$ within the super-cell approach, one has to find the corresponding index range $\mathcal{M}=\{m:|\theta_m-\theta_0|\leq\Delta\theta/2\}$, which depends on the period of the structure, so that $P_{sca}\approx\sum_{m\in\mathcal{M}}P_m$. One may be also interested in total reflected power, and has to sum over all propagating reflected plane waves taking $\theta_0=0$ and $\Delta\theta=\pi$. 

\section{Two approaches for analysis of semi-infinite structures}
\label{sec:twoapp}

Super-cell approach relying on the slicing of scattering structures and S-matrix calculation with the FMM can be exploited differently. Here we present two ways of calculating the reflectivity from random semi-infinite structures.
In the first approach, one considers a particular realization of a scattering structure.
The idea is to sequentially increase the thickness $H$ of a layer under consideration, starting with a top slice neighboring the above homogeneous half-space. Reflection matrix of the resulting layer is calculated through the reflection matrices of the individual thin slices combined together during the procedure. Convergence of $P_{sca}$ for an increasing number of slices under consideration would then provide an estimation of $P_{sca}|_{H\rightarrow\infty}$.

The second possible approach is analogous to the Ambartsumian's idea \cite{Ambartsumian1943} applied within the radiative transfer theory. Let us fix a finite scattering layer thickness $H=H_0$ being sufficient for this layer to represent all physical properties of the scattering medium under consideration. Then, due to the semi-infinite nature of the scattering structure, addition of such a layer on its top should not change the reflection properties. 
We also suppose the thickness $H_0$ to be large enough for the loss of coherence. Therefore, the idea 
should be formulated in terms of intensities of reflected plane waves. Then, instead of the amplitude matrix in Eq.~(\ref{eq:smat}) we take an analogous matrix relating the power carried by partial plane waves. Similar matrices are used when analyzing the light interaction with incoherent layers of media, such as thick glass substrates in experimental setups \cite{Centurioni2005Dec}. Reflection and transmission blocks of such matrices will be denoted with a bar to distinguish it from Eq.~(\ref{eq:smat}). We denote the unknown reflection matrix is $R_{aa}$, and the power S-matrix blocks describing the layer with an additional superscript '$L$'. Then the requirement for the latter matrix to remain the same upon application of the S-matrix multiplication rule yields the equation
\begin{equation}
	\overline{R}_{aa} = \overline{R}^{L}_{aa} + \overline{T}^{L}_{ab} \left( \mathbb{I} - \overline{R}_{aa} \overline{R}^{L}_{bb} \right)^{-1}\! \overline{R}_{aa} \overline{T}^{L}_{ba}
	\label{eq:rmat}
\end{equation}
Introducing $X=\left( \mathbb{I} - \overline{R}_{aa} \overline{R}^{L}_{bb} \right) (\overline{T}^{L}_{ab})^{-1}$ allows reducing the equation to the second order unilateral matrix Ricatti equation
\begin{equation}
	X^2A + XB + C = 0
	\label{eq:ric}
\end{equation}
with matrices $A = \overline{T}^{(L)}_{ab} (\overline{R}^{(L)}_{bb})^{-1}$, $B = \overline{R}^{(L)}_{aa} - (\overline{R}^{(L)}_{bb})^{-1} - \overline{T}^{(L)}_{ab} (\overline{R}^{(L)}_{bb})^{-1} \overline{T}^{(L)}_{ba}$, and $C = (\overline{R}^{(L)}_{bb})^{-1} \overline{T}^{(L)}_{ba}$. Considering eigenvalues $\lambda_i$ and eigenvectors $\bm{v}_i$ of matrix $X$, the attained quadratic equation can be reformulated as a double size generalized eigenvalue problem \cite{Higham2000}:
\begin{equation}
	\left(\! \begin{array}{cc} -C & 0 \\ 0 & \mathbb{I} \end{array} \!\right) \left(\! \begin{array}{c} \bm{v} \\ \bm{u} \end{array} \!\right) = 
	\lambda \left(\! \begin{array}{cc} B & \mathbb{I} \\ A & 0 \end{array} \!\right) \left(\! \begin{array}{c} \bm{v} \\ \bm{u} \end{array} \!\right)
	\label{eq:eig2}
\end{equation}
Assuming the size of matrices in Eq.~(\ref{eq:rmat}) to be $N_p$ the method yields $2N_p$ eigenvalues and eigenvectors. Providing that the initial amplitude S-matrix of each slice is calculated with sufficient accuracy, i.e. $N_h$ is sufficiently large, it appears that half of the eigenvalues are close to zero, and, therefore should be excluded. We verified that such a procedure leads to physically correct power matrices $\overline{R}_{aa}$ with all positive elements.


\section{Results and discussion}
\label{sec:res}

The two approaches described above appear to possess several numerical parameters affecting the resulting calculated reflectivity. At the lowest level an accuracy of solution of the diffraction problem for a single slice depends on the number of Fourier harmonics only. The S-matrix algorithm used for S-matrix calculation of structures with increasing thickness is well-known to be stable \cite{Cotter1995May}. Increasing the number of slices within the first approach leads to convergence of the reflectivity to a value being specific for a given sample of scattering medium. Further averaging over structure realizations (in the first approach applied to the entire structure, and in the second only to the additional layer)
yields averaged reflectivities for a fixed period. Finally, studying the dependence of the latter result on the structure period allows attaining an estimate of an averaged reflectivity for a semi-inifinite structure with given averaged optical and geometrical parameters. Here we discuss and demonstrate convergence at different levels of computations. The simplest example, which in principle can be benchmarked versus a 2D version of the MSTM method, is a homogeneous medium with constant permittivity filled with cylindrical inclusions. Additionally, a more realistic and complex example will be considered at the end of this section.

The convergence of the FMM for photonic crystal slab simulations is well-known to be polynomial \cite{Li2014}. One can formulate an empirical estimate of the number of Fourier harmonics $N_h$ sufficient to attain meaningful results for large periods:
\begin{equation}
	N_h\gtrsim \max\{4\Lambda/\lambda, 2\Lambda/\delta\}
	\label{eq:no_est}
\end{equation}
with $\delta$ being the smallest characteristic geometrical feature size on a period. Eq.~(\ref{eq:no_est}) means, that one has to consider inherently a sufficiently large number of evanescent waves while resolving small geometrical features in the reciprocal space.

\begin{figure}[h!]
    \includegraphics[width=0.5\textwidth]{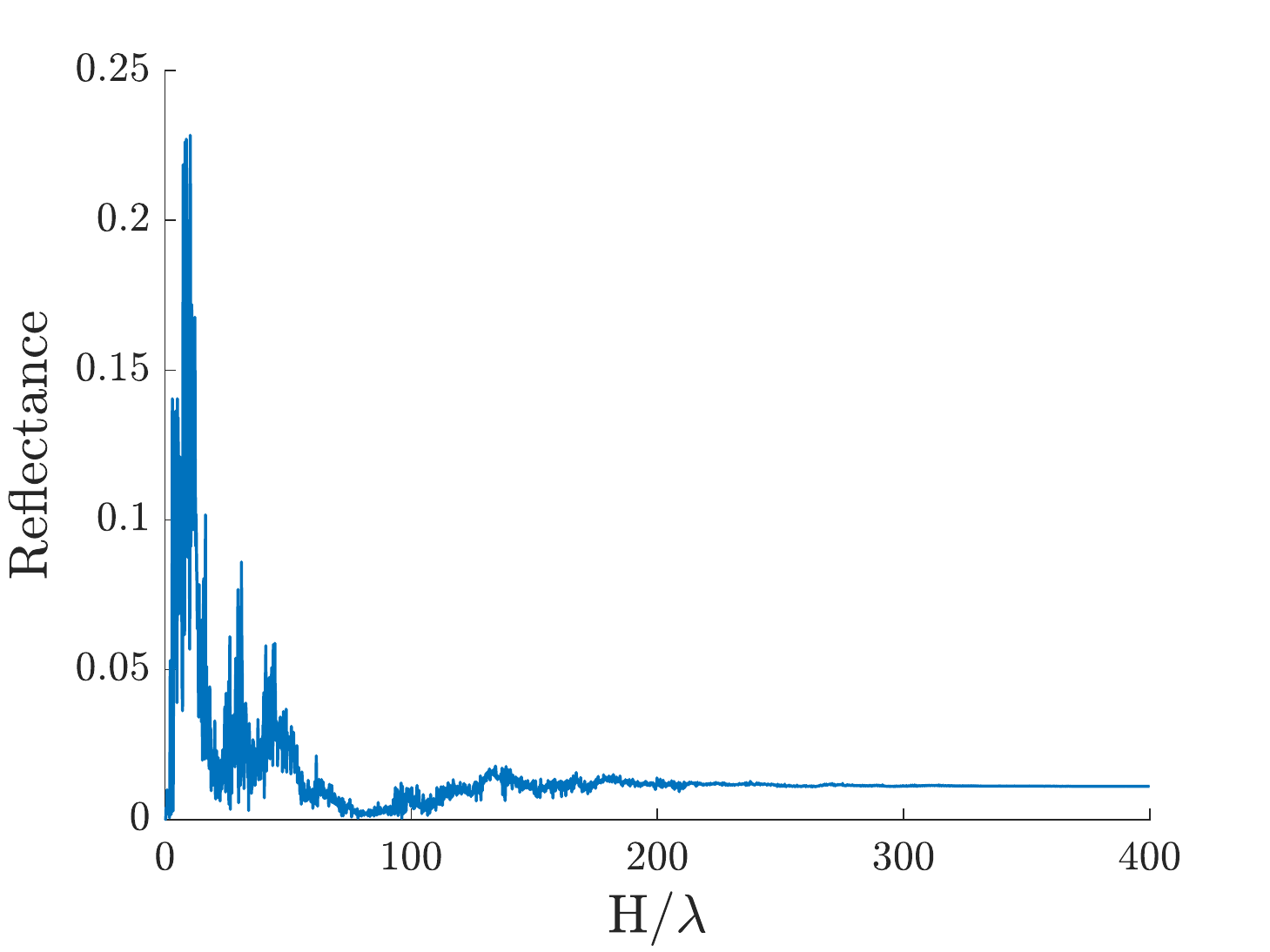}
    \caption{Dependence of the reflectance on the thickness of the structure for a given sample. Parameters of inclusions: radius of cylinders r = 0.4$\lambda$, concentration n = 0.3,  permittivity contrast $\Delta\varepsilon  = 1$. }
\label{fig:R(H)}
\end{figure}

\begin{figure}[h!]
    \includegraphics[width=0.5\textwidth]{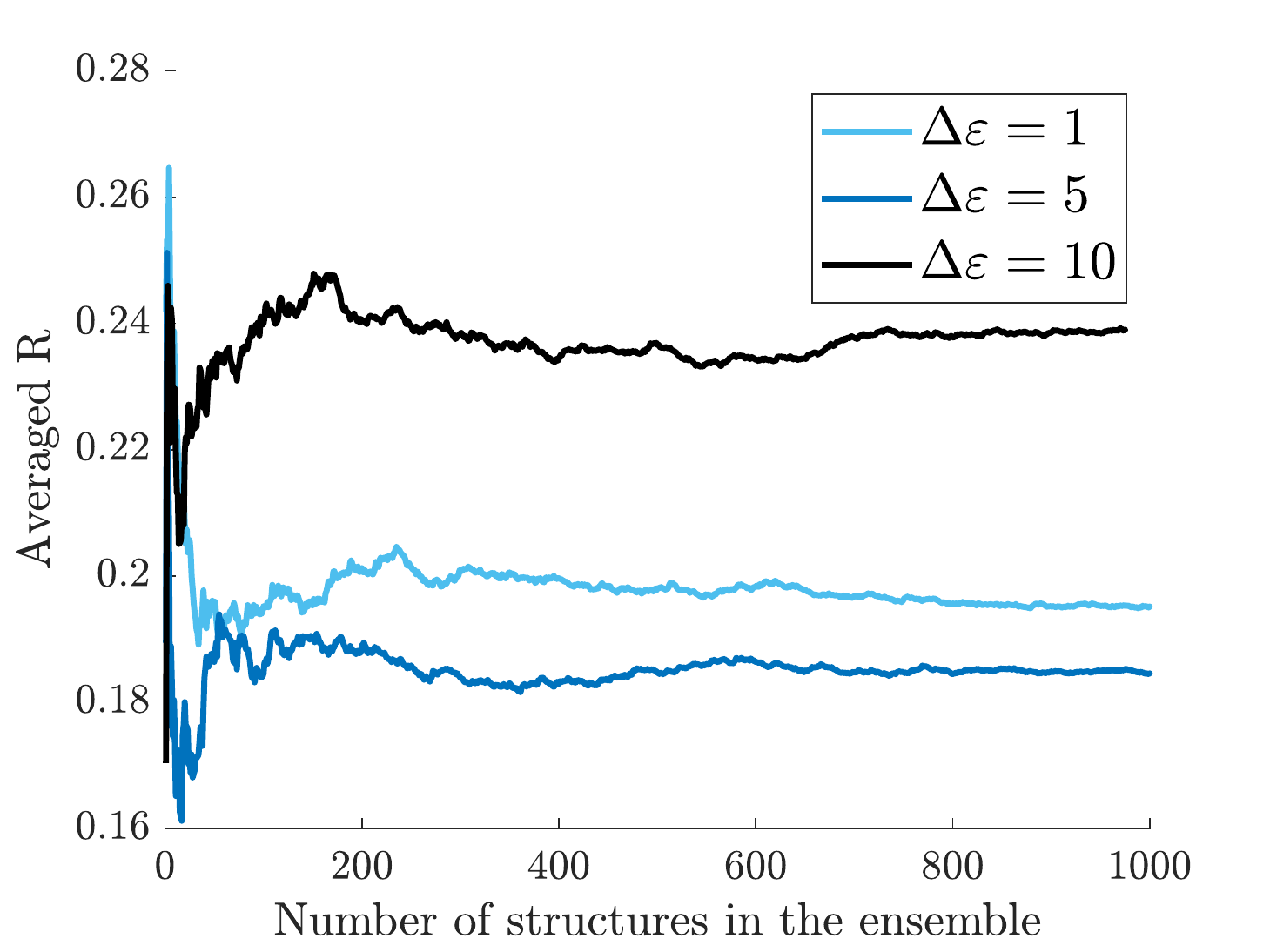}       
    \includegraphics[width=0.5\textwidth]{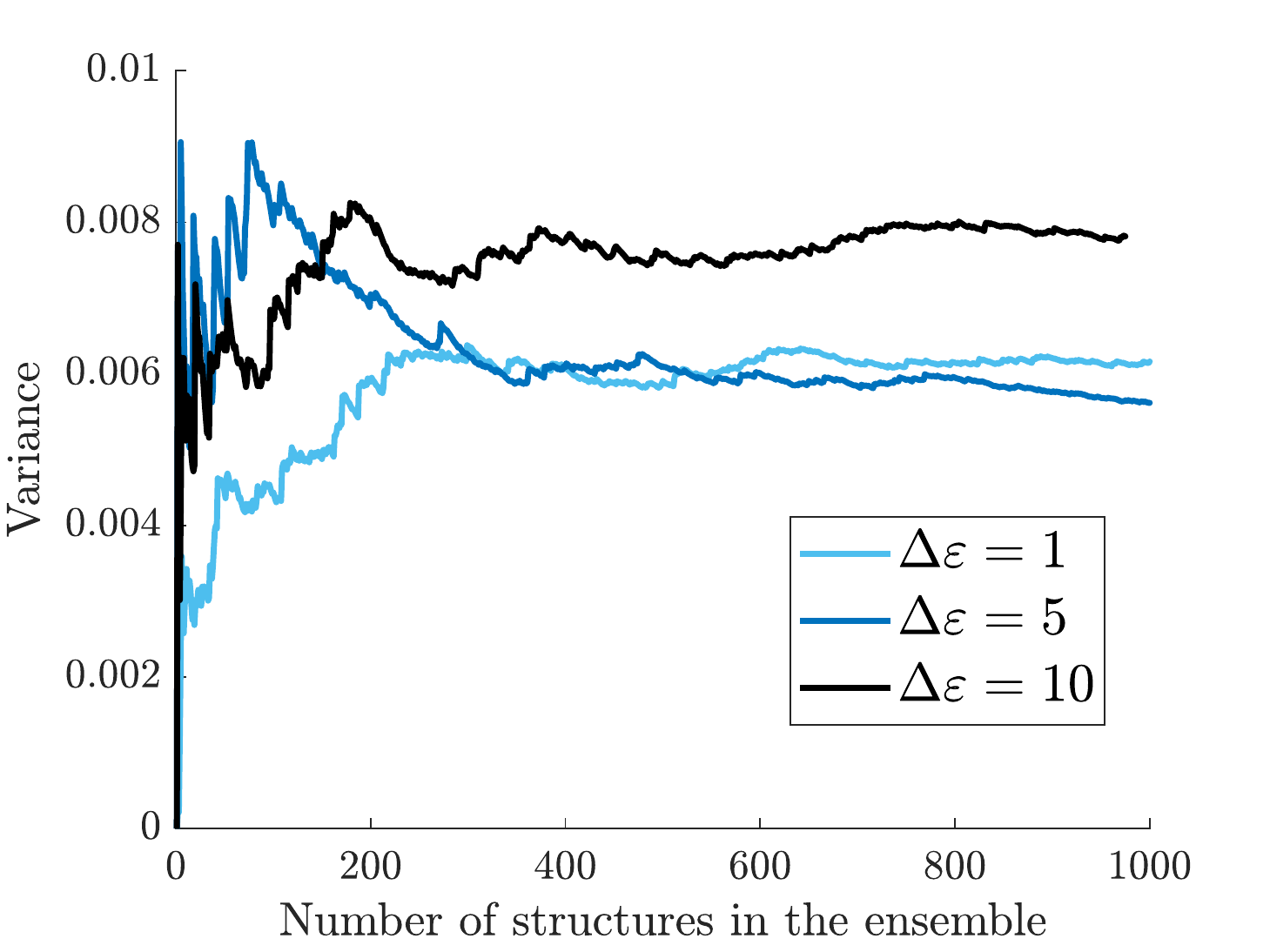}       
    \caption{(a) Averaged reflectivity, and (b) variance for different permittivity contrasts. Parameters of inclusions: radius of cylinders r = 0.4$\lambda$, concentration n = 0.3.}
\label{fig:R_av}
\end{figure}

Reflectivity of a scattering structure of cylindrical inclusions with a fixed super-cell period and increasing total thickness calculated within the first approach is demonstrated in Fig.~\ref{fig:R(H)}. It is seen that an effective penetration depth amounts to dozens of wavelengths and particle diameters in all considered examples. Further ensemble averaging yield dependencies of the reflectivity from ensemble number shown in Fig.~\ref{fig:R_av}. These dependencies allow us to define an ensemble averaged reflectivity $\overline{R}$ and corresponding variance
\begin{equation}
	\sigma^2 = \dfrac{\sum_{i=1}^{N}(R_i-\overline{R})}{N_e-1}
	\label{eq:var}
\end{equation}
with $N_e$ being the number of structure samples. 

\begin{figure}[h!]
    \includegraphics[width=0.5\textwidth]{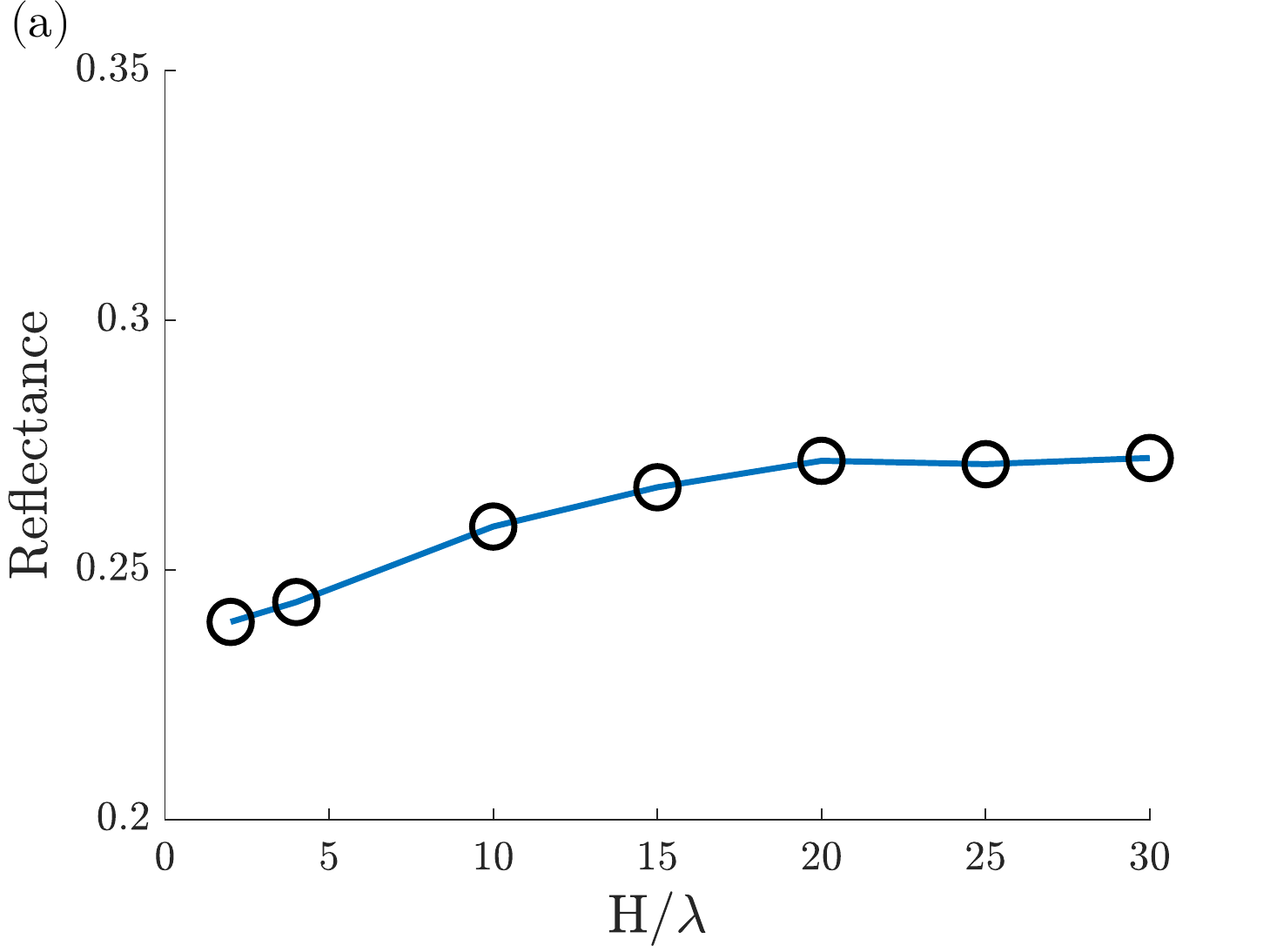}  
    \includegraphics[width=0.5\textwidth]{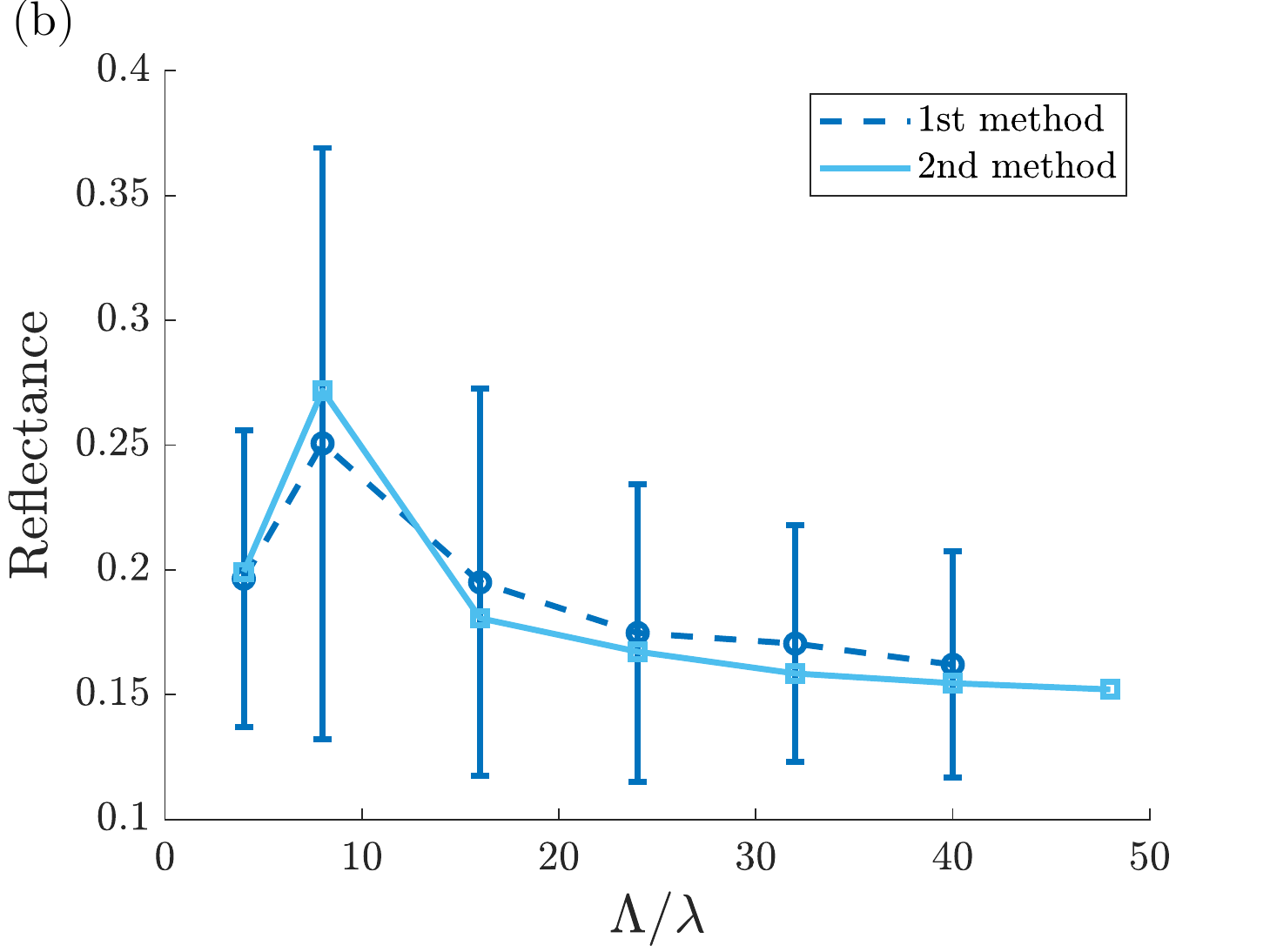} 
    \caption{Reflectivity calculated for the scattering structures with cylindrical inclusions. (a) Dependence of the reflectivity from the thickness of additional layer used within the second approach; (b) Dependencies of the reflectivity from the super-cell period attained within two approaches. }
\label{fig:R(period)_cyl}
\end{figure}

The second approach introduces an additional numerical parameter being the thickness $H_{0}$ of an averaged layer.  Fig.~\ref{fig:R(period)_cyl}(a) illustrates the dependence of $R(H)$. Interestingly, a sufficient thickness of this layer appears to be less, than the effective scattering penetration depth as seen from the results of the first approach.

Having attained the convergence for fixed super-cell periods ($\Lambda/\lambda$ in dimensionless units) we increased the period up to values when $\overline{R}$ becomes independent of $\Lambda$. This step required to consider the parameter $\Lambda/\lambda$ up to several dozens, as Fig.~\ref{fig:R(period)_cyl}(b) demonstrates. The final results approximate reflectivity of random semi-infinite scattering structures.

\begin{figure}[h!]
\centering
       \includegraphics[width=0.4\textwidth]{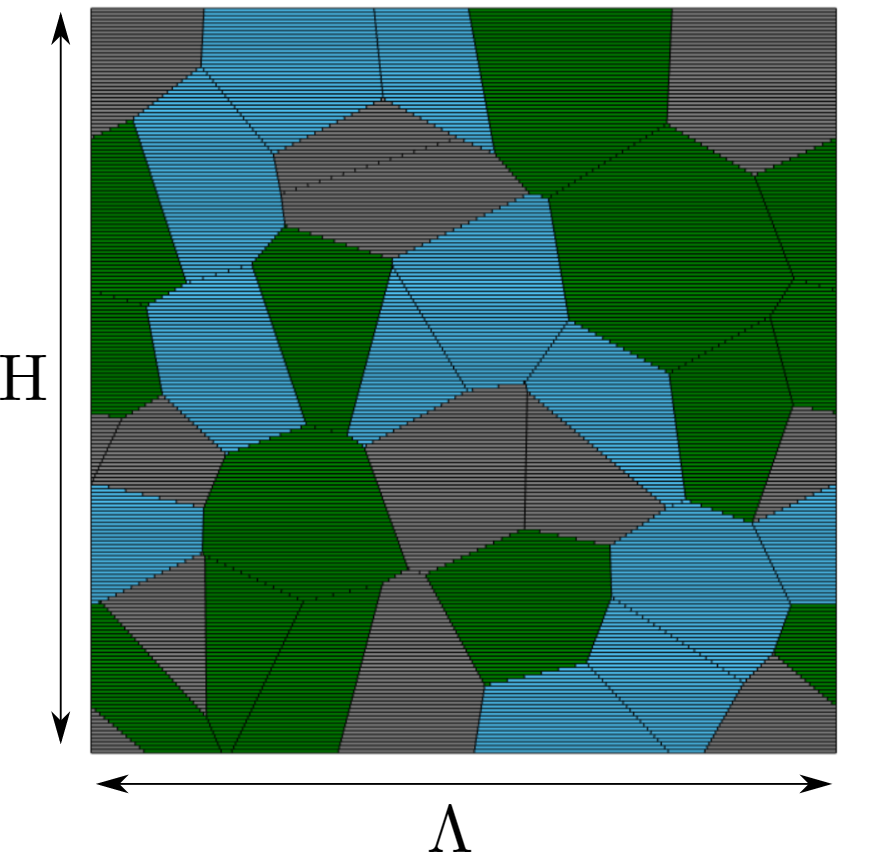}     
\caption{Example of a period of the scattering structure, wherein material boundaries a represented as a Voronoi tessellation. Horizontal lines indicate the slicing procedure.}
\label{fig:ill_voronoi}
\end{figure}

\begin{figure}[h!]
        \includegraphics[width=0.5\textwidth]{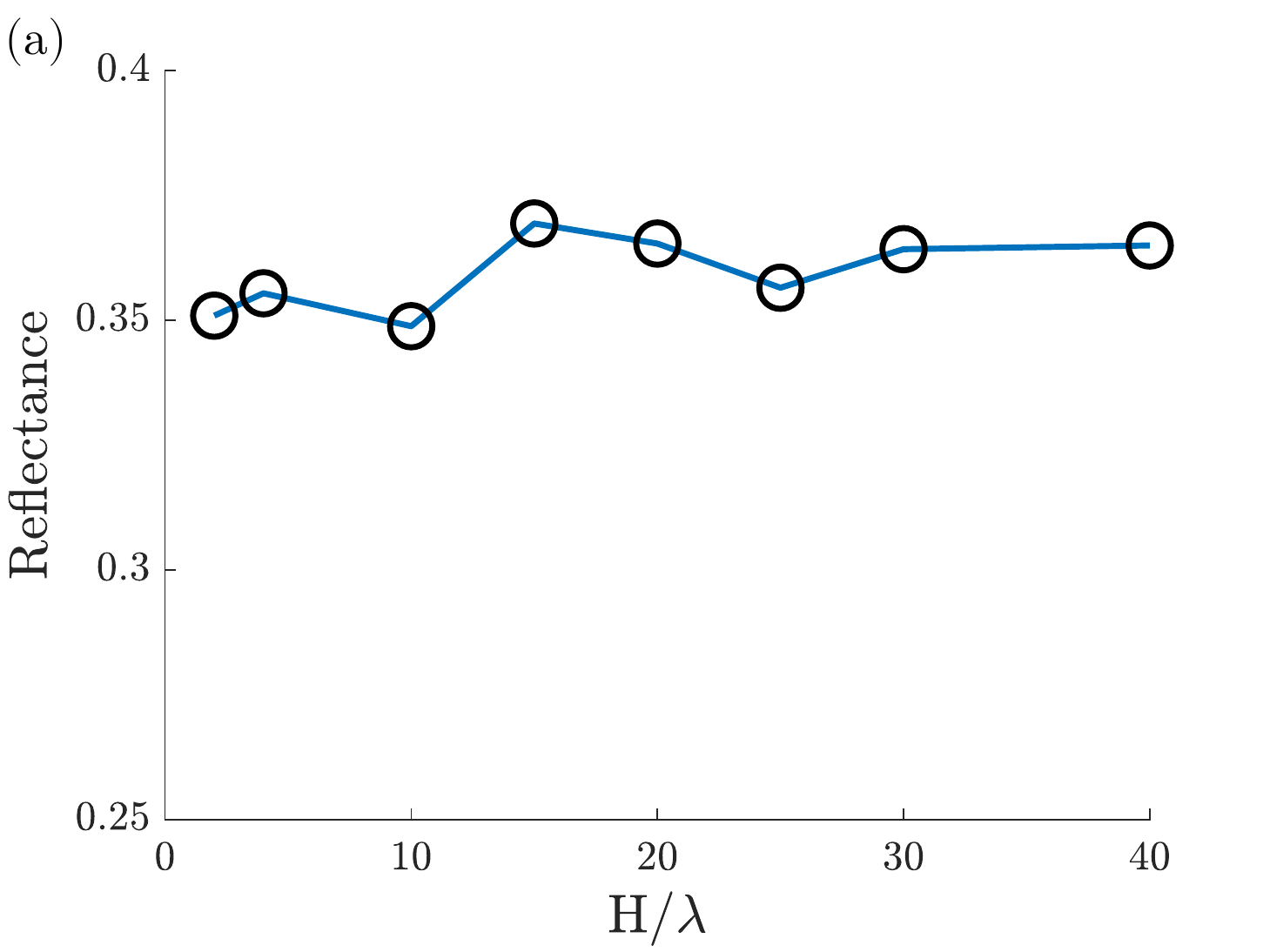}
         \includegraphics[width=0.5\textwidth]{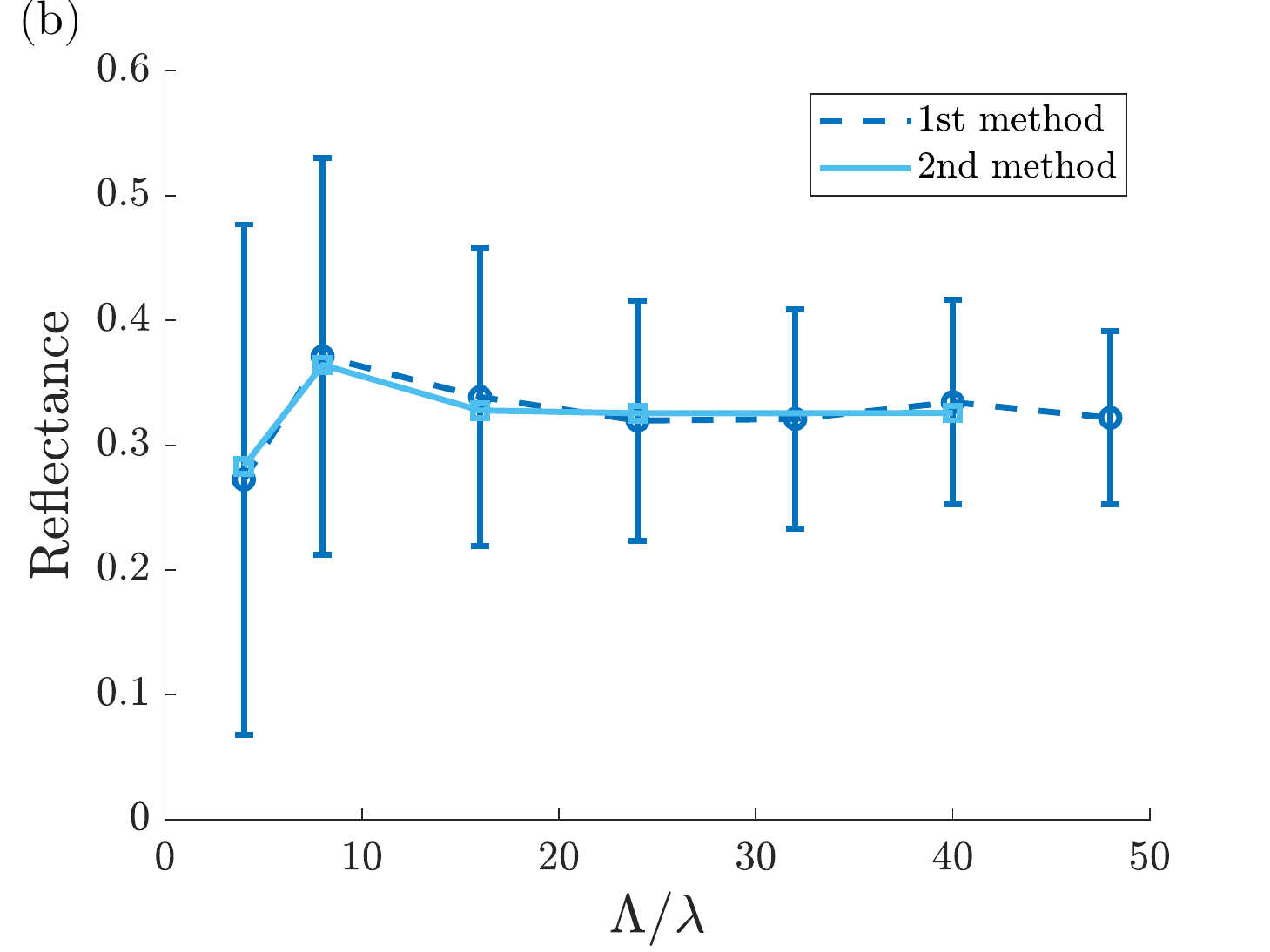}

\caption{Reflectivity calculated for the scattering structures generated with the Voronoi method. (a)
Dependence of the reflectivity from the thickness of additional layer used within the second approach; (b) Dependencies of the reflectivity from the super-cell period attained within two approaches. }
\label{fig:voronoi_results}
\end{figure}

In order to demonstrate the advantage of the discussed approaches over spherical/cylindrical harmonic decomposition based methods mentioned in the introduction we consider a structure with arbitrary polygonal boundaries separating materials with different permittivities. To generate such samples we used the Voronoi tessellation \cite{Aurenhammer2013}. An example of such a sample is demonstrated in Fig.~\ref{fig:ill_voronoi}, where different colors correspond to different materials, and the horizontal lines indicate slicing procedure. Convergence plots corresponding to such a structure are shown in Figs.~\ref{fig:voronoi_results} One can see, that these plots are essentially similar to the previous ones describing a medium of equal circular particles.

To summarize, in this work we have developed a method for calculation of the electromagnetic scattering by 2D semi-infinite strongly scattering media based on the super-cell approach and the Fourier Modal Method for grating diffraction calculation. Numerical examples demonstrate the consistency of the approach and its potential for analysis of 3D problem, which are inaccessible for other methods. The 3D case will definitely require invocation of specific numerical methods for efficient S-matrix computations, and which will be the subject of our future publication.

\section*{Acknowledgements}

This work was supported by the Russian Foundation for Basic Research, gr. No. 22-11-00153.



\section*{Appendix: Fourier Modal Method and S-matrix approach}
\label{app:fmm}

The mentioned FMM rationale consists in (a) solution of the eigenvalue problem for an infinite photonic crystal structure, (b) construction of a T-matrix of an interface between the semi-infinite photonic crystal and a homogeneous medium, and (c) construction of the S-matrix of a photonic crystal slab bounded by two planar interfaces with homogeneous substrate and superstrate media. In the simplest case of the collinear diffraction of the TE polarized plane wave the eigenvalue problem for the Fourier amplitude vector of the orthogonal electric field component $E_y$ reads
\begin{equation}
	\left( \omega^2\mu_{0} \hat{\varepsilon} - K^2  \right) \bm{E}_y = \beta^2 \bm{E}_y
	\label{eq:fmm_eig}
\end{equation}
Here $\hat{\varepsilon}$ is the Toeplitz matrix of periodic permittivity function Fourier amplitudes, $K={\rm{diag}}\{k_{mx}\}$ is the diagonal matrix of wavevector projections defined by the grating Eq.~(\ref{eq:grating}), and $\beta$ are mode propagation constants.

Once the eigenvectors $\bm{E}_{yn}$ and eigenvalues $\beta_n$ are found, the corresponding eigenvectors for the magnetic field $\bm{H}_{xn}$ are calculated from the Maxwell's equations. The eigenvectors consist of decomposition coefficients of the photonic crystal modes into plane waves. Given an interface between the photonic crystal and a homogeneous isotropic medium of permittivity $\varepsilon_0$ the T-matrix relates plane wave amplitudes $a_m^{\pm}$ and mode amplitudes $c_q^{\pm}$ in the following way:
\begin{equation}
\begin{split}
    \left(\!\!\begin{array}{c} a_m^+ \\ a_m^-  \end{array}\!\!\right) &= \sum_q T_{mq} \left(\!\!\begin{array}{c} c_q^+ \\ c_q^-  \end{array}\!\!\right) \\ &= \dfrac{1}{2}\sum_q \left(\!\!\begin{array}{cc} M^{-}_{mq} & M^{+}_{mq} \\ M^{+}_{mq} & M^{-}_{mq} \end{array}\!\!\right) \left(\!\!\begin{array}{c} c_q^+ \\ c_q^-  \end{array}\!\!\right)
\end{split}
\label{eq:fmm_tmat}
\end{equation}
with matrix elements $M^{\pm}_{mq} = E_{yqm}\pm(\omega\mu_0/k_{zm})H_{xqm}$. Finally, the S-matrix is written via blocks of substrate $T^{sub}$ and superstrate $T^{sup}$ T-matrices, and diagonal matrix $B=\mathrm{diag}\{\exp(i\beta_qh)\}$:
\begin{equation}
    S = Q_1Q_2^{-1},
    \label{eq:fmm_S}
\end{equation}
where
\begin{equation}
    Q_1 = \left(\!\!\begin{array}{cc} T^{sub}_{11} & T^{sub}_{12}B \\ T^{sup}_{21}B & T^{sup}_{22} \end{array}\!\!\right),
    \label{eq:fmm_Q1}
\end{equation}
and
\begin{equation}
    Q_2 = \left(\!\!\begin{array}{cc} T^{sub}_{21} & T^{sub}_{22}B \\ T^{sup}_{11}B & T^{sup}_{12} \end{array}\!\!\right),
    \label{eq:fmm_Q2}
\end{equation}

Owing the S-matrices of two adjacent photonic crystal slabs, $S^{(1)}$ and $S^{(2)}$, the S-matrix of a combined structure consisting of these two slabs is found from the composition rule, which is conventionally formulated for S-matrix blocks as
\begin{equation}
    R_{bb} = R_{bb}^{(1)} + T_{ba}^{(1)} \left( I - R_{bb}^{(2)} R_{aa}^{(1)} \right)^{-1} R_{bb}^{(2)} T_{ab}^{(1)}
    \label{eq:fmm_smmul_bb}
\end{equation}
\begin{equation}
    T_{ba} = T_{ba}^{(1)} \left( I - R_{bb}^{(2)} R_{aa}^{(1)} \right)^{-1} T_{ba}^{(2)}
    \label{eq:fmm_smmul_ba}
\end{equation}
\begin{equation}
    T_{ab} = T_{ab}^{(2)} \left( I - R_{aa}^{(1)} R_{bb}^{(2)} \right)^{-1} T_{ab}^{(1)}
    \label{eq:fmm_smmul_ab}
\end{equation}
\begin{equation}
    R_{aa} = R_{aa}^{(2)} + T_{ab}^{(2)} \left( I - R_{aa}^{(1)} R_{bb}^{(2)} \right)^{-1} R_{aa}^{(1)} T_{ba}^{(2)}
    \label{eq:fmm_smmul_aa}
\end{equation}

\printbibliography

@article{Tsang2017Jul,
	author = {Tsang, Leung and Liao, Tien-Hao and Tan, Shurun and Huang, Huanting and Qiao, Tai and Ding, Kung-Hau},
	title = {{Rough Surface and Volume Scattering of Soil Surfaces, Ocean Surfaces, Snow, and Vegetation Based on Numerical Maxwell Model of 3-D Simulations}},
	journal = {IEEE J. Sel. Top. Appl. Earth Obs. Remote Sens.},
	volume = {10},
	number = {11},
	pages = {4703--4720},
	year = {2017},
	month = {Jul},
	issn = {2151-1535},
	publisher = {IEEE},
	doi = {10.1109/JSTARS.2017.2722983}
}

@article{Penttila2021Mar,
	author = {Penttil{\ifmmode\ddot{a}\else\"{a}\fi}, Antti and Markkanen, Johannes and V{\ifmmode\ddot{a}\else\"{a}\fi}is{\ifmmode\ddot{a}\else\"{a}\fi}nen, Timo and R{\ifmmode\ddot{a}\else\"{a}\fi}bin{\ifmmode\ddot{a}\else\"{a}\fi}, Jukka and Yurkin, Maxim A. and Muinonen, Karri},
	title = {{How much is enough? The convergence of finite sample scattering properties to those of infinite media}},
	journal = {J. Quant. Spectrosc. Radiat. Transfer},
	volume = {262},
	pages = {107524},
	year = {2021},
	month = {Mar},
	issn = {0022-4073},
	publisher = {Pergamon},
	doi = {10.1016/j.jqsrt.2021.107524}
}

@article{Theobald2021Sep,
	author = {Theobald, Dominik and Beutel, Dominik and Borgmann, Luisa and Mescher, Henning and Gomard, Guillaume and Rockstuhl, Carsten and Lemmer, Uli},
	title = {{Simulation of light scattering in large, disordered nanostructures using a periodic T-matrix method}},
	journal = {J. Quant. Spectrosc. Radiat. Transfer},
	volume = {272},
	pages = {107802},
	year = {2021},
	month = {Sep},
	issn = {0022-4073},
	publisher = {Pergamon},
	doi = {10.1016/j.jqsrt.2021.107802}
}

@article{Tishkovets2006Sep,
	author = {Tishkovets, Victor P. and Jockers, Klaus},
	title = {{Multiple scattering of light by densely packed random media of spherical particles: Dense media vector radiative transfer equation}},
	journal = {J. Quant. Spectrosc. Radiat. Transfer},
	volume = {101},
	number = {1},
	pages = {54--72},
	year = {2006},
	month = {Sep},
	issn = {0022-4073},
	publisher = {Pergamon},
	doi = {10.1016/j.jqsrt.2005.10.001}
}

@article{Kahnert2016Jul,
	author = {Kahnert, Michael},
	title = {{Numerical solutions of the macroscopic Maxwell equations for scattering by non-spherical particles: A tutorial review}},
	journal = {J. Quant. Spectrosc. Radiat. Transfer},
	volume = {178},
	pages = {22--37},
	year = {2016},
	month = {Jul},
	issn = {0022-4073},
	publisher = {Pergamon},
	doi = {10.1016/j.jqsrt.2015.10.029}
}

@article{Gibson2005Feb,
	author = {Gibson, A. P. and Hebden, J. C. and Arridge, S. R.},
	title = {{Recent advances in diffuse optical imaging}},
	journal = {Phys. Med. Biol.},
	volume = {50},
	number = {4},
	pages = {R1-R43},
	year = {2005},
	month = {Feb},
	issn = {0031-9155},
	publisher = {IOP Publishing},
	doi = {10.1088/0031-9155/50/4/r01}
}

@article{Yu2021Mar,
	author = {Yu, Sunkyu and Qiu, Cheng-Wei and Chong, Yidong and Torquato, Salvatore and Park, Namkyoo},
	title = {{Engineered disorder in photonics}},
	journal = {Nat. Rev. Mater.},
	volume = {6},
	pages = {226--243},
	year = {2021},
	month = {Mar},
	issn = {2058-8437},
	publisher = {Nature Publishing Group},
	doi = {10.1038/s41578-020-00263-y}
}

@article{Centurioni2005Dec,
	author = {Centurioni, Emanuele},
	title = {{Generalized matrix method for calculation of internal light energy flux in mixed coherent and incoherent multilayers}},
	journal = {Appl. Opt.},
	volume = {44},
	number = {35},
	pages = {7532--7539},
	year = {2005},
	month = dec,
	issn = {2155-3165},
	publisher = {Optica Publishing Group},
	doi = {10.1364/AO.44.007532}
}

@article{Ambartsumian1943,
	author = {Ambartsumian, V. A. },
	title = {{On the question of diffuse reflection of light in turbid medium.}},
	journal = {Dokl. Akad. Nauk SSSR},
	volume = {38},
	number = {8},	
    pages = {257-261},
	year = {1943}
}

@article{Markkanen2017Mar,
	author = {Markkanen, J. and Yuffa, A. J.},
	title = {{Fast superposition T-matrix solution for clusters with arbitrarily-shaped constituent particles}},
	journal = {J. Quant. Spectrosc. Radiat. Transfer},
	volume = {189},
	pages = {181--188},
	year = {2017},
	month = mar,
	issn = {0022-4073},
	publisher = {Pergamon},
	doi = {10.1016/j.jqsrt.2016.11.004}
}

@article{Mackowski2011Sep,
	author = {Mackowski, D. W. and Mishchenko, M. I.},
	title = {{A multiple sphere T-matrix Fortran code for use on parallel computer clusters}},
	journal = {J. Quant. Spectrosc. Radiat. Transfer},
	volume = {112},
	number = {13},
	pages = {2182--2192},
	year = {2011},
	month = sep,
	issn = {0022-4073},
	publisher = {Pergamon},
	doi = {10.1016/j.jqsrt.2011.02.019}
}

@Article{Yoon2020Mar,
author={Yoon, S. and Kim, M. and Jang, M. and Choi, Y. and Choi, W. and Kang, S. and Choi, W.},
title={Deep optical imaging within complex scattering media},
journal={Nature Reviews Physics},
year={2020},
month={Mar},
day={01},
volume={2},
number={3},
pages={141-158},
issn={2522-5820},
doi={10.1038/s42254-019-0143-2},
url={https://doi.org/10.1038/s42254-019-0143-2}
}

@article{Jacucci2020Jul,
  doi = {10.1002/adma.202001215},
  url = {https://doi.org/10.1002/adma.202001215},
  year = {2020},
  month = jul,
  publisher = {Wiley},
  volume = {33},
  number = {28},
  pages = {2001215},
  author = {Gianni J. and Lukas S. and Yating Z. and Han Y. and Silvia V.},
  title = {Light Management with Natural Materials: From Whiteness to Transparency},
  journal = {Advanced Materials}
}

@inbook{Mishchenko2003,
  author={Mishchenko, M. I.},
  editor={van Tiggelen, B. A. and Skipetrov, S. E.},
  title={Radiative transfer theory: From Maxwell's equations to practical applications},
  booktitle={Wave Scattering in Complex Media: From Theory to Applications},
  year={2003},
  pages={367--414},
  publisher={Kluwer Academic},
  address={Dordrecht, the Netherlands},
}

@article{Roger2014,
title = {A hybrid transport-diffusion model for radiative transfer in absorbing and scattering media},
journal = {Journal of Computational Physics},
volume = {275},
pages = {346-362},
year = {2014},
issn = {0021-9991},
doi = {https://doi.org/10.1016/j.jcp.2014.06.063},
url = {https://www.sciencedirect.com/science/article/pii/S002199911400494X},
author = {Roger, M. and Caliot, C. and Crouseilles, N. and Coelho,  P.J.},
}

@article{Byrnes2022,
    author = {Byrnes, N. and Foreman, M. R.},
    title = {Random matrix theory of polarized light scattering in disordered media},
    journal = {Waves in Random and Complex Media},
    volume = {0},
    number = {0},
    pages = {1-29},
    year  = {2022},
    publisher = {Taylor & Francis},
    doi = {10.1080/17455030.2022.2153305},
    URL = {https://doi.org/10.1080/17455030.2022.2153305},
}

@article{Gushchin2010,
    author = {Gushchin, I. and Tishchenko, A. V.},
    journal = {J. Opt. Soc. Am. A},
    number = {7},
    pages = {1575--1583},
    publisher = {Optica Publishing Group},
    title = {Fourier modal method for relief gratings with oblique boundary conditions},
    volume = {27},
    month = {Jul},
    year = {2010},
    url = {https://opg.optica.org/josaa/abstract.cfm?URI=josaa-27-7-1575},
    doi = {10.1364/JOSAA.27.001575},
}

@article{Dovesi1994,
  doi = {10.1080/01411599408201207},
  url = {https://doi.org/10.1080/01411599408201207},
  year = {1994},
  month = nov,
  publisher = {Informa {UK} Limited},
  volume = {52},
  number = {2-3},
  pages = {151--167},
  author = {R. Dovesi and R. Orlando},
  title = {Convergence properties of the supercell approach in the study of local defects in solids},
  journal = {Phase Transitions}
}

@article{Elson2003,
  doi = {10.1088/0959-7174/13/2/303},
  url = {https://doi.org/10.1088/0959-7174/13/2/303},
  year = {2003},
  month = apr,
  publisher = {Informa {UK} Limited},
  volume = {13},
  number = {2},
  pages = {95--105},
  author = {J Merle Elson},
  title = {Scattering losses from planar waveguides with material inhomogeneity},
  journal = {Waves in Random Media}
}

@article{Jarvis1997,
  doi = {10.1103/physrevb.56.14972},
  url = {https://doi.org/10.1103/physrevb.56.14972},
  year = {1997},
  month = dec,
  publisher = {American Physical Society ({APS})},
  volume = {56},
  number = {23},
  pages = {14972--14978},
  author = {M. R. Jarvis and I. D. White and R. W. Godby and M. C. Payne},
  title = {Supercell technique for total-energy calculations of finite charged and polar systems},
  journal = {Physical Review B}
}

@article{Bulgakov2000,
  doi = {10.1088/0959-7174/10/3/302},
  url = {https://doi.org/10.1088/0959-7174/10/3/302},
  year = {2000},
  month = jul,
  publisher = {Informa {UK} Limited},
  volume = {10},
  number = {3},
  pages = {359--366},
  author = {S A Bulgakov and M Nieto-Vesperinas},
  title = {Defect-enhanced resonances in photonic lattices},
  journal = {Waves in Random Media}
}

@article{Ko1988,
  doi = {10.1364/josaa.5.001863},
  url = {https://doi.org/10.1364/josaa.5.001863},
  year = {1988},
  month = nov,
  publisher = {The Optical Society},
  volume = {5},
  number = {11},
  pages = {1863},
  author = {D. Y. K. Ko and J. R. Sambles},
  title = {Scattering matrix method for propagation of radiation in stratified media: attenuated total reflection studies of liquid crystals},
  journal = {Journal of the Optical Society of America A}
}

@article{Higham2000,
  doi = {10.1093/imanum/20.4.499},
  url = {https://doi.org/10.1093/imanum/20.4.499},
  year = {2000},
  month = oct,
  publisher = {Oxford University Press ({OUP})},
  volume = {20},
  number = {4},
  pages = {499--519},
  author = {N. J. Higham},
  title = {Numerical analysis of a quadratic matrix equation},
  journal = {{IMA} Journal of Numerical Analysis}
}

@book{Aurenhammer2013,
  title     = {Voronoi Diagrams and Delaunay Triangulations},
  author    = {Aurenhammer, F. and Klein, R. and Lee, D.-T.},
  year      = {2013},
  publisher = {World Scientificy},
}

@inbook{Li2014,
  author={Li, L.},
  editor={Popov, E.},
  title={Fourier Modal Method},
  booktitle={ Gratings: Theory and Numeric Applications, Second
Revisited Edition},
  year={2014},
  pages={13.1--13.40},
  publisher={Aix Marseille Université},
  %address={Dordrecht, the Netherlands},
}

@article{Cotter1995May,
	author = {Cotter, N. P. K. and Preist, T. W. and Sambles, J. R.},
	title = {{Scattering-matrix approach to multilayer diffraction}},
	journal = {J. Opt. Soc. Am. A, JOSAA},
	volume = {12},
	number = {5},
	pages = {1097--1103},
	year = {1995},
	month = may,
	issn = {1520-8532},
	publisher = {Optica Publishing Group},
	doi = {10.1364/JOSAA.12.001097}
}

\end{document}